\documentclass[12pt,preprint]{aastex}
\usepackage{amsmath}
\usepackage{amssymb}
\usepackage{xspace}
\usepackage{color}
\usepackage{xspace}
\usepackage{natbib}

\newcommand{\powersep}{{\ensuremath{\times}}}

\newcommand{\g}{{\ensuremath{\mathrm{g}}}\xspace}
\newcommand{\K}{{\ensuremath{\mathrm{K}}}\xspace}
\newcommand{\cm}{{\ensuremath{\mathrm{cm}}}\xspace}

\newcommand{\km}{{\ensuremath{\mathrm{km}}}\xspace}
\newcommand{\Msun}{{\ensuremath{\mathrm{M}_{\odot}}}\xspace}
\newcommand{\Sec}{{\ensuremath{\mathrm{s}}}\xspace}
\newcommand{\erg}{{\ensuremath{\mathrm{erg}}}\xspace}
\newcommand{\ergs}{{\ensuremath{\erg\,\Sec^{-1}}}\xspace}

\newcommand{\kms}{{\ensuremath{\km\,\Sec^{-1}}}\xspace}
\newcommand{\cms}{{\ensuremath{\cm\,\Sec^{-1}}}\xspace}
\newcommand{\foe}{{\ensuremath{\Ep{51}\,\erg}}\xspace}
\newcommand{\Foe}{{\ensuremath{\powersep\foe}}\xspace}

\newcommand{\kB}{{\ensuremath{\mathrm{k}_{\mathrm{B}}}}\xspace}
\newcommand{\NA}{{\ensuremath{\mathrm{N}_{\mathrm{\!A}}}}\xspace}
\newcommand{\cL}{{\ensuremath{\mathrm{c}}}\xspace}

\newcommand{\opaunit}{{\ensuremath{\cm^2\,\g^{-1}}}\xspace}
\newcommand{\Deg}{{\ensuremath{^\circ}}\xspace}

\newcommand{\lSect}[1]{{\label{sec:#1}}}
\newcommand{\lFig}[1]{{\label{fig:#1}}}

\newcommand{\lTab}[1]{{\label{tab:#1}}}

\newcommand{\Tabff}[1]{{\ref{tab:#1}}}
\newcommand{\Tab}[1]{{Table~\Tabff{#1}}}

\newcommand{\pan}[1]{{\textit{#1}}}

\newcommand{\FIGFF}[2]{{\ref{fig:#2}\pan{#1}}}

\newcommand{\FIG}[2]{{Fig.~\FIGFF{#1}{#2}}}
\newcommand{\Fig}[1]{{\FIG{}{#1}}}

\newcommand{\Sectff}[1]{{\ref{sec:#1}}}
\newcommand{\Sect}[1]{{\S~\Sectff{#1}}}

\newcommand{\isofont}[1]{{\mathrm{#1}}}
\newcommand{\isomass}[1]{{\ensuremath{\isofont{^{#1}}}}}
\newcommand{\isocharge}[1]{{\ensuremath{\isofont{_{#1}}}}}
\newcommand{\isotope}[3]{{\ensuremath{\isocharge{#1}\isomass{#2}\isofont{#3}}}}

\newcommand{\I}[2]{{\isotope{}{#1}{#2}}}

\newcommand{\Ep}[1]{{\ensuremath{10^{#1}}}}

\newcommand{\E}[1]{{\ensuremath{\powersep\Ep{#1}}}}
\newcommand{\EE}[2]{{\ensuremath{\powersep\Ep{#1#2}}}}

\slugcomment{Draft for ApJ, \today}

\shorttitle{THE LIGHT CURVE OF THE UNUSUAL \ SUPERNOVA 2003DH}
\shorttitle{Woosley, \& Heger}

\begin{document}

\title{THE LIGHT CURVE OF THE UNUSUAL SUPERNOVA \ 2003DH}
\author{S. E. Woosley}
\affil{
  Department of Astronomy and Astrophysics, 
  University of California,
  Santa Cruz, CA 95064
}
\email{woosley@ucolick.org}

\and

\author{A. Heger}
\affil{
   Theoretical Astrophysics Group,
   T-6, MS B227,
   Los Alamos Nation Laboratory,
   Los Alamos, NM 87545, 
and\\ 
   Enrico Fermi Institute,
   University of Chicago,
   5640 S.\ Ellis Ave.,
   Chicago, IL 60637
}
\email{1@2sn.org}

\begin{abstract}
SN 2003dh, one of the most luminous supernovae ever recorded, and the
one with the highest measured velocities, accompanied gamma-ray burst
030329. Its rapid rise to maximum and equally rapid decline pose
problems for any spherically symmetric model. We model the supernova
here as a very energetic, polar explosion that left the equatorial
portions of the star almost intact. The total progenitor mass was much
greater than the mass of high-velocity ejecta, and the total mass of
\I{56}{Ni} synthesized was about 0.5 solar masses. Such asymmetries
and nickel masses are expected in the collapsar model.  A ``composite
two-dimensional'' model is calculated that agrees well with the
characteristics of the observed light curve. The mass of \I{56}{Ni}
required for this light curve is $0.55\,\Msun$ and the total explosion
energy, $26\,\Foe$.

\end{abstract}


\section{INTRODUCTION}

On March 29, 2003 one of the brightest gamma-ray bursts (GRBs) in
history was discovered and localized by the HETE-2
satellite. Astronomers worldwide watched and within days were rewarded
by the discovery of a Type Ic supernova, SN 2003dh, in precisely the
same location \citep{Sta03,Hjo03}. From the light curve and spectrum,
a temporal coincidence with the burst was also estimated to be
$\lesssim 2$ days.  It is now generally agreed that the supernova and
the GRB came from the same explosion.

The V-band light curve and spectrum of SN 2003dh closely resembled
that of another famous supernova-GRB pair, SN 1998bw \citep{Gal98} and
GRB 980425, but with several important distinctions: \textbf{1)} GRB
980425 had an observed equivalent isotropic energy in gamma-rays
roughly four orders of magnitude less than GRB 030329; \textbf{2)} SN
2003dh rose to maximum in less than 10 days, SN 1998bw took 16 days
\citep{Woo99}; \textbf{3)} SN 2003dh exhibited higher expansion
speeds, up to $40,000\,\kms$; and \textbf{4)} the total energy of
ejected relativistic matter in SN 1998bw was $\lesssim 3\EE50$ with $v
> 0.5\,\cL$ \citep{Li99}. This is less even than the energy in
gamma-rays from GRB 030329, unless the beaming angle is very small.

It is generally agreed that SN 1998bw was a very asymmetric explosion
whose high velocities may not have characterized ejecta at all angles
\citep[e.g.,][]{Maz01,Mae03}, and the same seems likely to be true of SN
2003dh. Despite the fact that its radiation is not beamed like a GRB,
an asymmetric supernova can have a quite different light curve from a
spherical one of the same total energy.  The higher velocities in one
direction lead to the earlier escape of radiation and, in the case of
a radioactive power source, a rapid decline in the deposition
efficiency of gamma-rays.  If the radioactivity itself is mixed out
preferentially along one axis, the efficiency of gamma-deposition is
also affected, leading to a more rapid rise and decline of the
luminosity.  On the other hand, the less rapidly moving matter ejected
in other directions, can continue to contribute an extended tail on the
light curve, powered by the remaining radioactivity.

In a pioneering study of light curves from asymmetric explosions,
\citet{Hof99} modeled the light curve of SN 1998bw, but attempted to
fit it into the general family of Type Ib/c supernovae. Their
parameters were thus typical of these common events: \I{56}{Ni} mass
($0.07$ to $0.2\,\Msun$), ejected mass ($2\,\Msun$), and explosion
energy ($2\,\Foe$).  They, and \citet{Woo99}, championed
the idea that asymmetrically exploded supernovae of the same energy
may have different light curves. However, in light of SN 2003dh,
neither went far enough.

Here we model the supernova as essentially two components - a slowly
moving, high-mass equatorial ejection and a lower mass polar ejection
\citep[see also][]{Mae02,Mae03}.  The juncture between these is almost
discontinuous.  A smooth transition would give too broad a light
curve.  The total mass of the progenitor, $\sim10\,\Msun$ of helium
and heavy elements, the \I{56}{Ni} mass synthesized, $\sim0.5\,\Msun$,
and the total explosion energy, $\sim \Ep{52}\,\erg$, are all quite
atypical for ordinary Type Ib/c supernovae.

\section{EXPLOSION MODELS}

A series of explosions was calculated for two Wolf-Rayet (WR) stars of
final mass $8.39\,\Msun$ and $15\,\Msun$.  Both models had an initial
mass of $15\,\Msun$ and an initial composition appropriate to the
helium core of a star with 0.1 solar metallicity.  Both models were
started with a surface rotation rate corresponding to $30\,\%$
Keplerian at the equator. It was assumed, however, that the lower mass
star was in a binary system and lost its hydrogen envelope to the
companion star during the expansion phase after core hydrogen
depletion (Case B mass transfer). This model continued to lose mass as
a WR-star at a rate given by \citet{Bra97}, reduced by
$\sqrt{[\mathrm{Fe}/\mathrm{Fe}_{\odot}]}$ (i.e., about a factor 3),
and another factor 3 to account for clumping \citep{HK98}.  The higher
mass model, while lacking a hydrogen envelope, assumed no further mass
loss.  This was an artificial way to create two rapidly rotating
WR-stars with a range of masses.  For the given starting mass, the one
evolved with mass loss is probably the more realistic.

In both calculations the effects of rotationally induced mixing were
included \citep{Heg00} throughout the evolution and both reached iron
core collapse with sufficient angular momentum to form a Kerr black
hole and an accretion disk.  The adopted reaction rate for the
\I{12}C($\alpha$,$\gamma$)\I{16}O reaction was 1.2 times that of
\citet{Buc96}.  Both models had very fine surface zoning, down to less
than $10^{21}\,\g$.  The radii of the two stars at the time their
cores collapsed were $8.1\EE10\,\cm$ (with mass loss) and
$8.8\EE10\,\cm$ (without mass loss).  Some other properties of the
presupernova stars are given in Tables \ref{tab:proj1} and
\ref{tab:proj2}.

Explosions were simulated by placing a piston at the location of a
large entropy jump, around $S/\NA\kB = 4$, in the presupernova star when
its peak infall velocity had reached about $1000\,\kms$.  Such a large
change in entropy typically corresponds to a sudden decrease in
density at the base of the oxygen-burning shell where mass
bifurcations often develop in supernova models.  The star outside this
piston was first allowed to collapse to $500\,\km$ at one-fourth the
free fall acceleration (inward movement of the piston). 
For the lower mass model, which only formed a small iron core
(\Tab{proj2}), the entropy jump and piston were at $1.462\,\Msun$.
For the higher mass model, a large collapsing low-entropy core formed
(\Tab{proj2}) and two piston locations were explored. One was located
at 1.93 $\Msun$, at the edge of the iron core, the other at
$2.75\,\Msun$ where the large entropy discontinuity occurred.  

After the minimum radius was reached, the piston was moved outward
supersonically, decelerating at constant fraction of gravitational
acceleration until it coasted to a halt at $10,000\,\km$.  The
initial velocity and deceleration of the piston were adjusted to give
the desired explosion energy.  (The right piston acceleration has been
determined using a modified regular FALSI algorithm.)  This explosion
energy here is defined as the final kinetic energy of the ejecta for
an explosion into vacuum.

A wide range of kinetic energies was explored for these isotropic
explosion (\Tab{mods}), including energies all the way up to
$1.6\E{53}\,\erg$.  This energy, half the binding energy of a typical
neutron star, is far more than expected from any realistic
neutrino-powered model, but, as we shall see, might be relevant for a
small amount of mass ejected in a very asymmetric explosion in which
the energy source is not neutrinos, but gravitational energy from
accretion into a black hole. Nucleosynthesis was calculated as
described in \citet{Wea78}.  An interesting result was the observation
of a maximum mass of \I{56}{Ni}, regardless of explosion energy and
depending only on the mass of the progenitor and depth of the piston.

A near constant mass, $0.2\pm0.05\,\Msun$ was synthesized in the
$8.39\,\Msun$ models (\Tab{mods}).  The near constancy of this upper
bound is a consequence of the fast expansion and increasing entropy in
the center of the ejecta that leads to freeze-out of \I4{He} rather
then \I{56}{Ni} (\Fig{comp8}).  Eventually, turning up the energy only
increases the ejection of $\alpha$-particles, not of \I{56}{Ni}.
Because the location of the piston cannot be much deeper in the
$8.39\,\Msun$ model and because any other way of exploding the star,
say by energy deposition rather than a piston, would give more
photodisintegration, $\sim0.2\,\Msun$ is the maximum \I{56}{Ni} that
can be synthesized in a spherically symmetric explosion of this star.
A larger amount can be made in the $15\,\Msun$ star without mass loss
because more mass sits closer to the piston.  For a piston mass of
$2.75\,\Msun$, the limiting mass of \I{56}{Ni} was $\sim0.6\,\Msun$.
For a piston mass of $1.93\,\Msun$, the edge of the iron core, the
limiting \I{56}{Ni} mass was $1.1\,\Msun$ (\Tab{mods}). One might get
still larger \I{56}{Ni} masses by going to helium stars above 15
$\Msun$. However, this is the largest helium core one expects for
stars near solar metallicity and larger mass cores will require even
greater energies to expand rapidly enough to explain SN 2003dh.

That these limits are within a factor of two of the mass of \I{56}{Ni}
inferred (see below) for both SN 1998bw and SN 2003dh is interesting,
but probably coincidental. The existence of an upper bound for shock
powered models has interesting implications though (\Sect{disc}).

\section{MODEL LIGHT CURVES}

Light curves were calculated for all the explosion models using the
\textsc{Kepler} code as described, e.g., in \citet{Woo99}. Gamma-rays
from \I{56}{Ni} and \I{56}{Co} decay were assumed to deposit
locally. The gamma-ray opacity was $0.037\,\opaunit$.  For the
diffusing radiation, opacity was assumed to be predominantly electron
scattering.  The electron density was calculated by solving the Saha
equation for the ionization structure at each point. Though calculated
assuming a single-temperature, flux-limited diffusion, and a simple
model for gamma-ray deposition, the curves should be qualitatively
correct and suffice for present purposes.

\Fig{lite15} and the first frame of \Fig{lite8} show the light curves
expected when the supernova experiences ``moderate'' mixing.  Mixing
was simulated by a running average of the composition within a region
of mass, $\Delta m$, set here to 10\,\% of the final mass of the star.
That is, the actual composition gradients were smoothed by averaging
within a defined band of masses, and this band was itself moved out,
zone by zone, from the piston to the surface.  This procedure was
repeated four times in each model.  For the 8.38\,\Msun model with
1.25\,\Foe explosion energy, for example, this resulted in \I{56}{Ni}
being mixed far enough out that it had 50\,\% of its central value
(i.e., just above the piston) at about $3\,\Msun$ and 10\,\% at
$5.5\,\Msun$ (as measured from the center of the collapsed
remnant). The second panel of \Fig{lite8} shows similar light curves
for the $8.35\,\Msun$ model when the composition is completely
homogenized, i.e., made to be the same from center top surface. This
might be the case if a vigorous asymmetric flow is responsible for
exploding the star.

These are bolometric light curves, which properly should be compared
only with luminosities integrated across the UVOIR bands. However,
based upon our experience with SN 1998bw, we shall compare them with
the observed V-band light curve of SN 2003dh \citep{Hjo03}. This shows
the brightness rising to maximum at 10 -- 13 days (rest frame) and
declining by 1 magnitude when the supernova was about 30 days old.
The authors further estimate that the supernova was ``slightly
brighter'' than SN 1998bw for which a peak luminosity
$1.0\E{43}\,\ergs$ \citep{Woo99} and \I{56}{Ni} mass $0.3$ to
$0.4\,\Msun$ has been estimated \citep{Nak01,Sol02}.

Though the \I{56}{Ni} mass ejected varies with kinetic energy and
progenitor mass (\Tab{mods}), the supernova must have made a single
value. Knowing the approximate luminosity we needed at peak, another
series of light curves was calculated for which the yield of
\I{56}{Ni} was normalized to $0.5\,\Msun$ in all the explosions, even
for the low mass progenitor. This is a typical value for the very
energetic explosions (Table 2), is close to the $0.3\,\Msun$
\citep{Sol02} to $0.4\,\Msun$ \citep{Nak01} of \I{56}{Ni} inferred for
SN 1998bw, and is consistent with the observation that SN 2003dh was a
little brighter than SN 1998bw \citep{Hjo03}. The fact that it is
allowed to exceed the hydrodynamical limit derived in the previous
section is justified because of the alternative way \I{56}{Ni} is made
in the collapsar model (\Sect{2d}).

Such energetic models with so much \I{56}{Ni} are likely to be
thoroughly mixed, more so than usual supernovae.  Mixing proved
necessary to obtain a good fit to the light curve of SN 1998bw, which
accompanied GRB 980425 \citep{Chu00}, and is favored by the rapid rise
time in SN 2003dh. The third panel in \Fig{lite8} shows the light
curves expected for the $8.39\,\Msun$ models when the ejecta are
thoroughly mixed and forced to contain a fiducial $0.5\,\Msun$ of
\I{56}{Ni}.  This was accomplished by replacing as much of the
material directly above the piston by pure \I{56}{Ni} as was needed to
obtain the desired total \I{56}{Ni} mass. The replacement was done
100\,\Sec after the explosion when essentially all thermonuclear
reactions have ceased, but before any mixing was applied.

\Fig{vfinal} shows the terminal velocity for both models as a function
of mass and \Fig{vphot} gives the velocity at the photosphere as a
function of time and for the $8.38\,\Msun$ model. In \Fig{vfinal} only
the intrinsically produced \I{56}{Ni} was considered, however, the
energy release from \I{56}{Ni} decay has little effect on the
expansion velocities, especially for the cases with high explosion
energy.  The fact that speeds as high as $\log\left(v/\cms\right) =
9.5$ were seen on SN 2003dh on day 10 \citep{Hjo03} illustrates the
need for at least some portion of the ejecta to have equivalent
isotropic energies above $4\E{52}\,\erg$. However, most of the models
can give the milder requirement of $\log\left(v
/\cms\right)\approx9.0$ on day $30$.

\section{A ``TWO-DIMENSIONAL'' MODEL}
\lSect{2d}

Only the most energetic symmetric explosions, $E\gtrsim\Ep{53}\,\erg$,
rise rapidly enough, make sufficient \I{56}{Ni}, and decay rapidly
enough to resemble SN 2003dh. The velocities in these models are also
consistent with what was seen (\Fig{vphot}), but the energies strain
credibility and even the most energetic model would not make a
powerful gamma-ray burst.  The correct model must be asymmetric.

Ideally, one would like to explore the coupled GRB-supernova explosion
in two-dimensional code that couples radiation transport, gamma-ray
deposition, and special relativistic hydrodynamics.  Such calculations
will surely be done, but as an expedient and for clarity in
exposition, we consider here a simple, composite toy model rendered
out of combinations of our one-dimensional models.

We rely here on experience with the collapsar model, especially
\citet{Mac99} and \citet{Zha02}. A black hole forms in the middle of
the star and, after a few seconds while the polar accretion rate
declines, launches relativistic jets along both axes. These jets
penetrate the star and ultimately make the GRB, but they do not, by
themselves, make a supernova. The peak brightness of a Type I
supernova of any subclass is measured by the \I{56}{Ni} that it
ejects. Both the jets and the lateral shocks they launch make almost
none. \emph{Lacking additional sources of power, the supernova
accompanying a collapsar-produced GRB would be weak and nearly
invisible}.

That additional source of power is the wind off the accretion disk
\citep{Mac99,Mac02,Pru03}.  The mass ejected is of the same order as
the mass accreted by the black hole, i.e., $\sim 1\,\Msun$. It's
energy is uncertain, but could easily be $\sim \Ep{52}\,\erg$, that is
the binding energy of a solar mass of material at the last stable
orbit around a Kerr black hole times a few percent. A similar amount
of energy is released by the reassembly of $0.5\,\Msun$ of nucleons
from the disk into bound nuclei.  \emph{Indeed the energy in this
``wind'' may considerably exceed the energy in the GRB-producing jets
themselves}. Its composition is likely to be chiefly \I{56}{Ni}
\citep{Pru03} mixed with the helium, oxygen, and other heavy elements
that make up the star.

This wind drives a highly asymmetric explosion \citep{Mac99}. In fact,
to first order, it blows two conical-shaped ``wedges'' out of the
star, each along the rotational axis. In the case of SN 2003dh,
because we saw the GRB, one of these inverted cones was directed
straight at us. The equatorial regions are partly ejected, but partly
fall into the hole, continuing to power the jet for some time after
the initial explosion.  The opening angle of the conical wedges is
unknown, but certainly greater than the $\sim5\Deg$ opening of the GRB
jet itself, yet probably small enough to contain only a fraction of
the stellar mass.  Here we will use $45\Deg$ as an example. This
implies that $1-\cos\theta = 29\,\%$ of the star is ejected at very
high velocity.  Within this $45\Deg$ we assume a Gaussian distribution
of kinetic energies between $160\,\Foe$ and $40\,\Foe$ (\Fig{eiso}) as
a function of angle.  We further assume a total production of
$0.55\,\Msun$ of \I{56}{Ni}, $90\,\%$ of which comes out in the high
velocity (well-mixed) wedges, and $10\,\%$ of which stays behind in
the low velocity ejecta.

The composite light curves were computed by assuming an explosion
energy as a function of colatitude, $\theta$:
\begin{equation}
E(\theta)=\left\{
\begin{array}{ll}
\phantom{.}160\,\Foe \times \exp\left\{-0.5\left(\theta/\theta_1\right)^{\!2}\right\} & \mbox{for}\quad\theta \le \pi/4 \\
1.25\,\Foe & \mbox{for}\quad\theta > \pi/4 \\
\end{array}
\right.
\end{equation}
where $\theta_1$ is chosen such that $E(\pi/4)=40\,\Foe$, i.e.,
\begin{equation}
\theta_1=\frac{\pi}{4}\left(2\log\left(\frac{160\,\Foe}{\phantom{0}40\,\Foe}\right)\right)^{\!\!-1/2}
\;.
\end{equation}
This was used to determine the flux as function of $\theta$ by
interpolation in the grid of one-dimensional light curves. The
one-dimensional models were calculated with an effective mass of
\I{56}{Ni} equal to $1.5\,\Msun$ (last panel of \Fig{lite8}), i.e., a
mass such that $29\,\%$ of it gave the actual \I{56}{Ni} mass in the
high-velocity ejecta, $0.44\,\Msun$.

These contributions were then integrated over the projection of the
sphere in the direction of the observer, along the pole (\Fig{slice}).
Such a procedure does not take into account the ``shadowing'' effect
of the fast moving polar ejecta on the slower equatorial ejecta.
Owing to its fast expansion, the polar ejecta quickly become optically
thin and the late-time light curve from the equatorial ejecta should
not be significantly affected.  We also do not take into account that
the cone of high explosion energy may trap $\gamma$-rays less
efficiently than a full sphere and thus cool down faster, or
similarly, that the polar ``hole'' punched into the explosion by the
jet could allow photons to escape faster in this direction, or that a
larger fraction of the low-energy ejecta than corresponds to the
projected surface area would become visible once the polar ejecta are
optically thin.  These truly two-dimensional effects may make the
low-E part of the ejecta a little brighter early on and decline a bit
faster at late times before the spherical model becomes optically
thin.

The composite light curve corresponding to these assumptions is given
in \Fig{composite}. During the first month, the luminosity is given
entirely by the high-velocity, nickel-rich ejecta. Because of the high
velocity and complete mixing the rise to maximum is very rapid.
Because the material becomes thin and gamma-rays from radioactive
decay escape, the decline from maximum is also fast.  The decline of
the light curve might be even faster than suggested by the composite
model, because $\gamma$-rays can escape easier from a polar wedge than
from a full sphere that has been used for modeling the light curves.

At late times though, the slower moving, nickel-poor ejecta dominate
the light curve because gamma-ray deposition there remains highly
efficient.  For situations where the energy is due chiefly to
radioactivity, this sort of light curve is distinctive signature of
asymmetry. Models that expand slowly would not be so bright at peak
and models that expand rapidly would not be so bright at late times.
The amount of \I{56}{Ni} inferred from the peak brightness would be
quite different from that inferred on the tail. Such seems to have
been the case for SN 1998bw where a \I{56}{Ni} mass of $0.7\,\Msun$
was inferred for the peak \citep{Iwa98} and $0.24\,\Msun$ at late
times \citep{Nak01}.

\section{DISCUSSION}
\lSect{disc}

The optical light curve of the supernova that accompanies a typical
GRB is unique in two ways. First, it must be disentangled from the
``optical afterglow'', which is dominated at early times by the
relativistic GRB-producing jet interacting with the circum-source
medium. There will also be optical emission from slower moving, but
still highly energetic supernova ejecta running into this
medium. Continued, possibly time-variable output from the central
engine also contributes at a declining rate to the afterglow at all
wavelengths \citep{Zha02}. Whether one counts these afterglows as
``supernova'' or something else is largely a matter of
taste. Supernova shock interactions have often been considered part of
the light curve \citep{Lei94}. They are probably the origin of
variations in the early optical emission seen during the first few
days of SN 2003dh \citep{Wil03}. Even with complete mixing, enormous
explosion energy, and a large mass of \I{56}{Ni}, a supernova powered
by radioactivity would not reach maximum light in two days.

Second, the supernova is grossly asymmetric. We have attempted to
account for this by merging the results of several spherically
symmetric models of various energies. Even so, large energies are
required in the composite model, far above the GRB jet energy
estimated by \citet{Fra01}. This energy must be provided by another
source which, in the collapsar model, is the disk wind. Any attempt to
model SN 2003dh without the formation of a black hole and disk must
find a different way to produce the necessary \I{56}{Ni} and energy.
The only other potential candidate at this time is a millisecond
magnetar \citep{Whe00}. Provided the putative pulsar can survive
accretion in a massive progenitor star long enough, it will need to
deposit $\sim\Ep{53}\,\erg$ in $\sim1\,\Sec$ in order to make
$\sim0.5\,\Msun$ of \I{56}{Ni} in a high mass progenitor (making this
much in a low mass model is impossible).  The \I{56}{Ni} is required
by the light curve and the high energy is required to make the star
move fast enough to explain the rapid post-maximum decline of the
supernova. The one-second time scale is set by the requirement that
the \I{56}{Ni} be made by a shock that raises the temperature to
greater than $5\E9\,\K$ as is necessary to produce \I{56}{Ni} out of
lighter elements.  Concentrating the pulsar's energy in a smaller
solid angle, as has been done here, will give a smaller \I{56}{Ni}
mass no greater than $1.1\,\Msun$ times that solid angle
(\Tab{mods}). In addition, the pulsar would need to focus some part of
this energy into a very relativistic narrow jet (in order to make the
GRB) at a time when the accretion rate is very high.

These potential difficulties lead us to favor the collapsar model.
Based upon this model, in which the \I{56}{Ni} is provided by the disk
wind in an explosion that is highly asymmetric we have developed a
composite working model in which the total explosion energy is
$26\,\Foe$ and the mass of \I{56}{Ni} is $0.55\,\Msun$.  Further
observations to test the predicted light curve (\Fig{composite}) are
encouraged.

\acknowledgments

This research has been supported by NASA (NAG5-8128, NAG5-12036, and
MIT-292701) and the DOE Program for Scientific Discovery through
Advanced Computing (SciDAC; DE-FC02-01ER41176).  AH has been supported
in part by the Department of Energy under grant B341495 to the Center
for Astrophysical Thermonuclear Flashes at the University of Chicago.


\clearpage
\begin{deluxetable}{cccc}
\tablecaption{Properties\tablenotemark{a} \ \ of the $15\,\Msun$
progenitor model without mass loss\lTab{proj1}}
\tablewidth{0pt}
\tablehead{
\colhead{} &
\colhead{$m$} &
\colhead{$r$} &
\colhead{$J$} 
\\
\colhead{} &
\colhead{($\Msun$)} &
\colhead{($\cm$)} &
\colhead{($\erg\ $s)}
}
\startdata
iron core      & $ 1.95$ & $2.58\E{ 8}$ & $1.23\E{50}$ \\
Si core        & $ 2.61$ & $4.99\E{ 8}$ & $2.35\E{50}$ \\
Ne/Mg/O core   & $ 2.95$ & $7.06\E{ 8}$ & $2.68\E{50}$ \\
C/O core       & $ 8.56$ & $5.16\E{ 9}$ & $2.24\E{51}$ \\
star / He core & $15.00$ & $8.80\E{10}$ & $1.00\E{52}$ \\
\tablenotetext{a}{Enclosed mass, $m$, radius, $r$, and enclosed angular
momentum, $J$, of the progenitor model at core collapse or the outer
boundaries of the indicated cores.}
\enddata
\end{deluxetable}

\clearpage
\begin{deluxetable}{cccc}
\tablecaption{Properties of the $8.38\,\Msun$ progenitor model with
mass loss\lTab{proj2}}
\tablewidth{0pt}
\tablehead{
\colhead{} &
\colhead{$m$} &
\colhead{$r$} &
\colhead{$J$} 
\\
\colhead{} &
\colhead{($\Msun$)} &
\colhead{($\cm$)} &
\colhead{($\erg\ $s)}
}
\startdata
iron core      & $ 1.46$ & $1.63\E{ 8}$ & $4.20\E{49}$ \\
Si core        & $ 1.77$ & $5.06\E{ 8}$ & $5.57\E{49}$ \\
Ne/Mg/O core   & $ 1.88$ & $6.77\E{ 8}$ & $5.96\E{49}$ \\
C/O core       & $ 4.71$ & $6.12\E{ 9}$ & $3.50\E{50}$ \\
star / He core & $ 8.38$ & $8.09\E{10}$ & $1.38\E{51}$ \\
\enddata
\end{deluxetable}

\clearpage

\begin{deluxetable}{cccc}
\tablecaption{Parameters of the Explosions \lTab{mods}}
\tablewidth{0pt}
\tablehead{
\colhead{Mass ($\Msun$)} &
\colhead{8.38} &
\colhead{15} &
\colhead{15} 
\\
\colhead{Piston ($\Msun$)} &
\colhead{1.45} &
\colhead{1.93} &
\colhead{2.75}
}
\startdata
KE (\foe) & & \I{56}{Ni} ($\Msun$) & \\
1.25 & 0.158 & 0.040\tablenotemark{a} & 0.024\tablenotemark{a} \\
2.5  & 0.193 & 0.614\tablenotemark{a} & 0.225                  \\
 5   & 0.229 & 0.707                  & 0.254                  \\
10   & 0.244 & 0.815                  & 0.303                  \\
20   & 0.230 & 0.938                  & 0.377                  \\
40   & 0.189 & 1.045                  & 0.487                  \\
80   & 0.168 & 1.083                  & 0.583                  \\
160  & 0.170 & 0.955                  & 0.625                  \\
\tablenotetext{a}{The \I{56}{Ni} mass is reduced by fallback
after mild mixing.  For the piston location at $1.93\,\Msun$
the \I{56}{Ni} masses before fallback are $0.558\,\Msun$ and
$0.623\,\Msun$; the total fallback mass is $6.889\,\Msun$ and
$0.050\,\Msun$ (mass coordinates $8.819\,\Msun$ and $1.980\,\Msun$).
For the piston location at $2.75\,\Msun$ the \I{56}{Ni} mass without
fallback is $0.192\,\Msun$, the fallback mass is $5.580\,\Msun$ (mass
coordinate $8.330\,\Msun$).}
\enddata
\end{deluxetable}

\clearpage
\begin{figure} \centering
\includegraphics[angle=0,height=0.4\textheight]{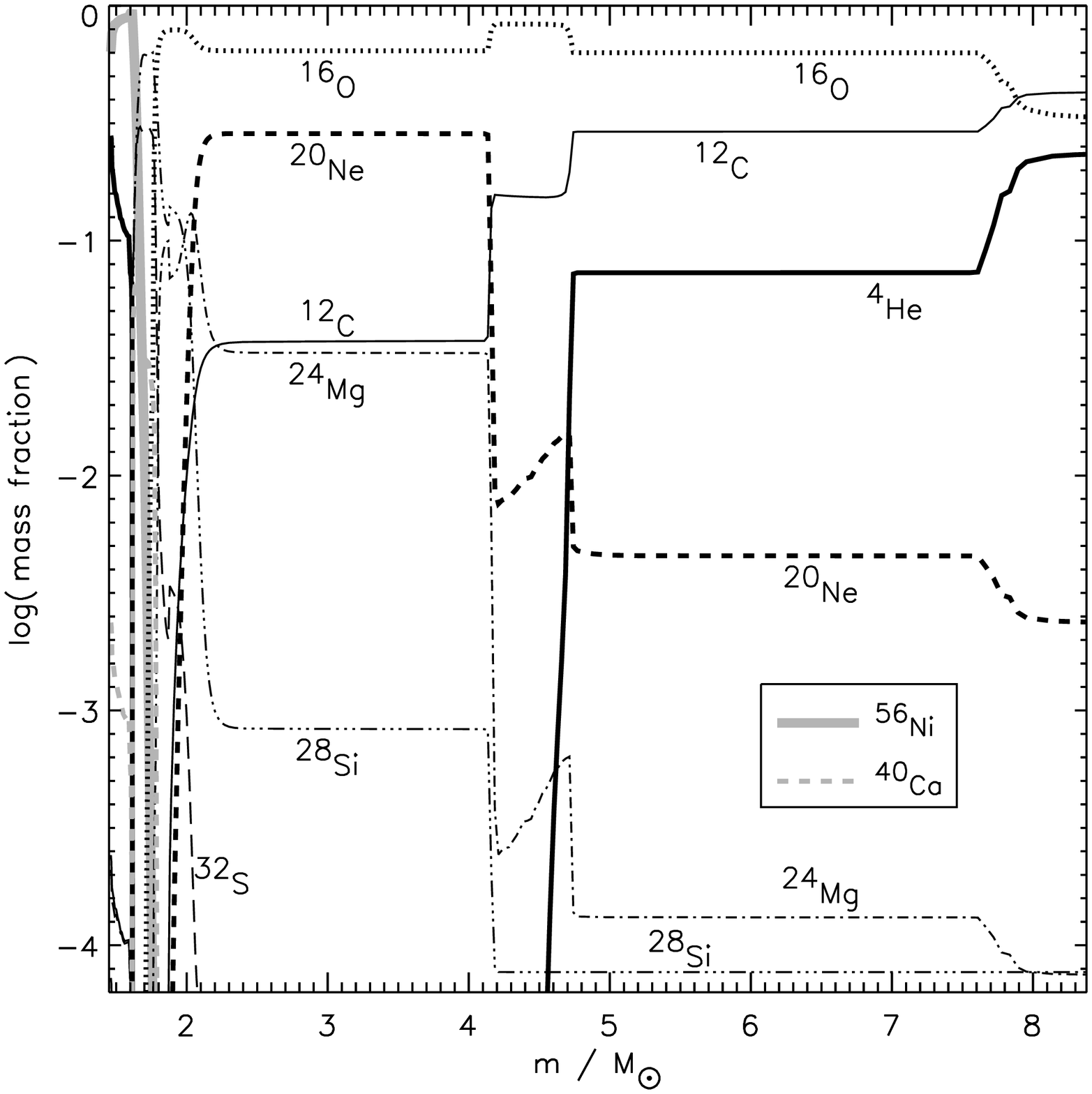}
\includegraphics[angle=0,bb=69 112 580 678,clip,height=0.4\textheight]{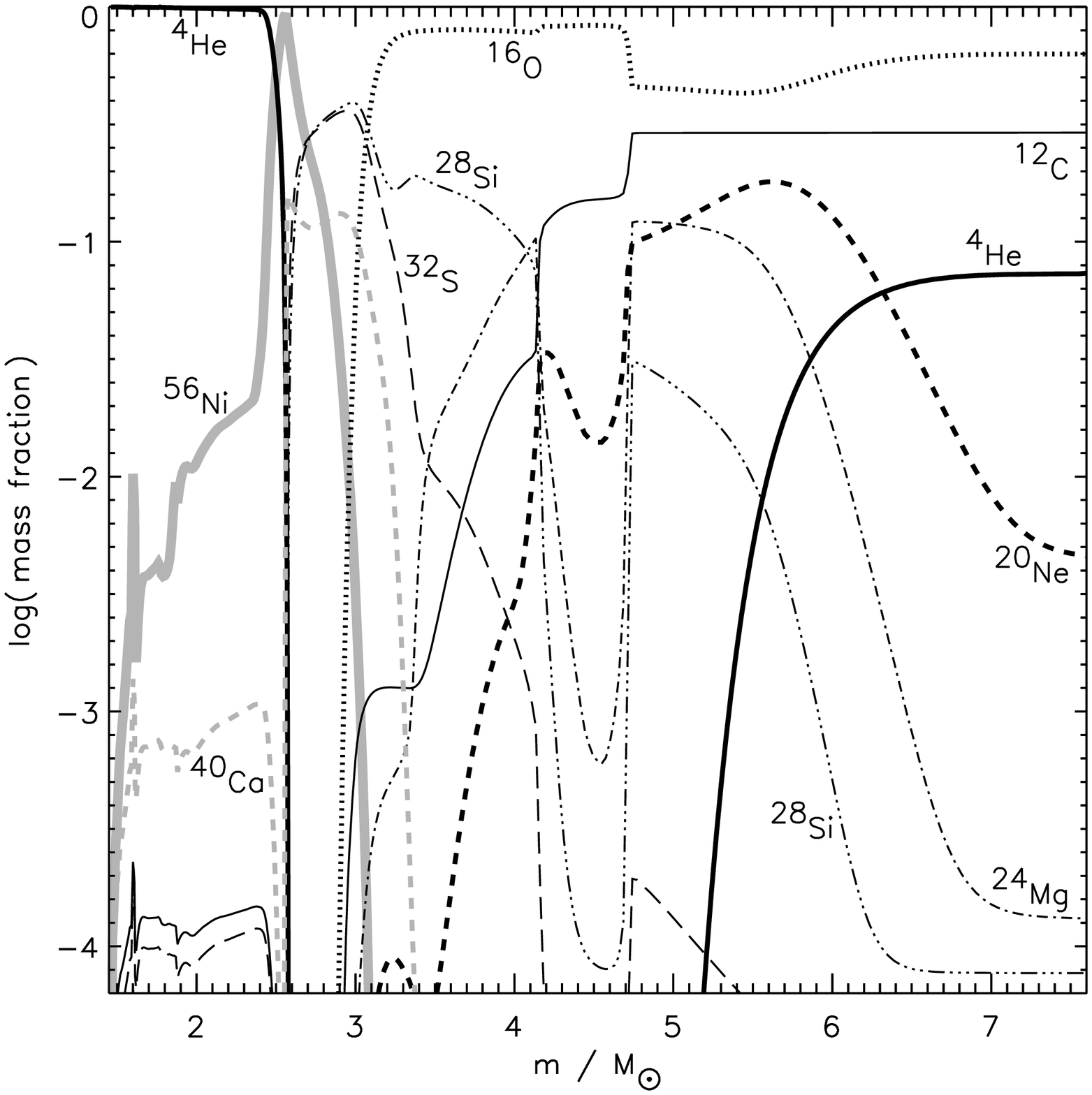}
\caption{\emph{Left:} Composition of the $8.38\,\Msun$ star with $1.25\,\Foe$
explosion energy $100\,\Sec$ after core collapse, but prior to any
mixing.  \emph{Right:} Same plot, but for the $\ 160\, \Foe$ explosion. Note 
that $0.8\,\Msun$ has been removed from the surface of this
calculation because it exceeded $\Ep{10}\,\cms$ and posed problems
for the non-relativistic hydro-code. Also note the large mass of
photodisintegrated matter (\I4{He}) in the inner regions near the mass
cut. This limits the production of \I{56}{Ni} in the model. \lFig{comp8}}
\end{figure}

\clearpage
\begin{figure} \centering
\includegraphics[angle=90,height=0.6\textheight]{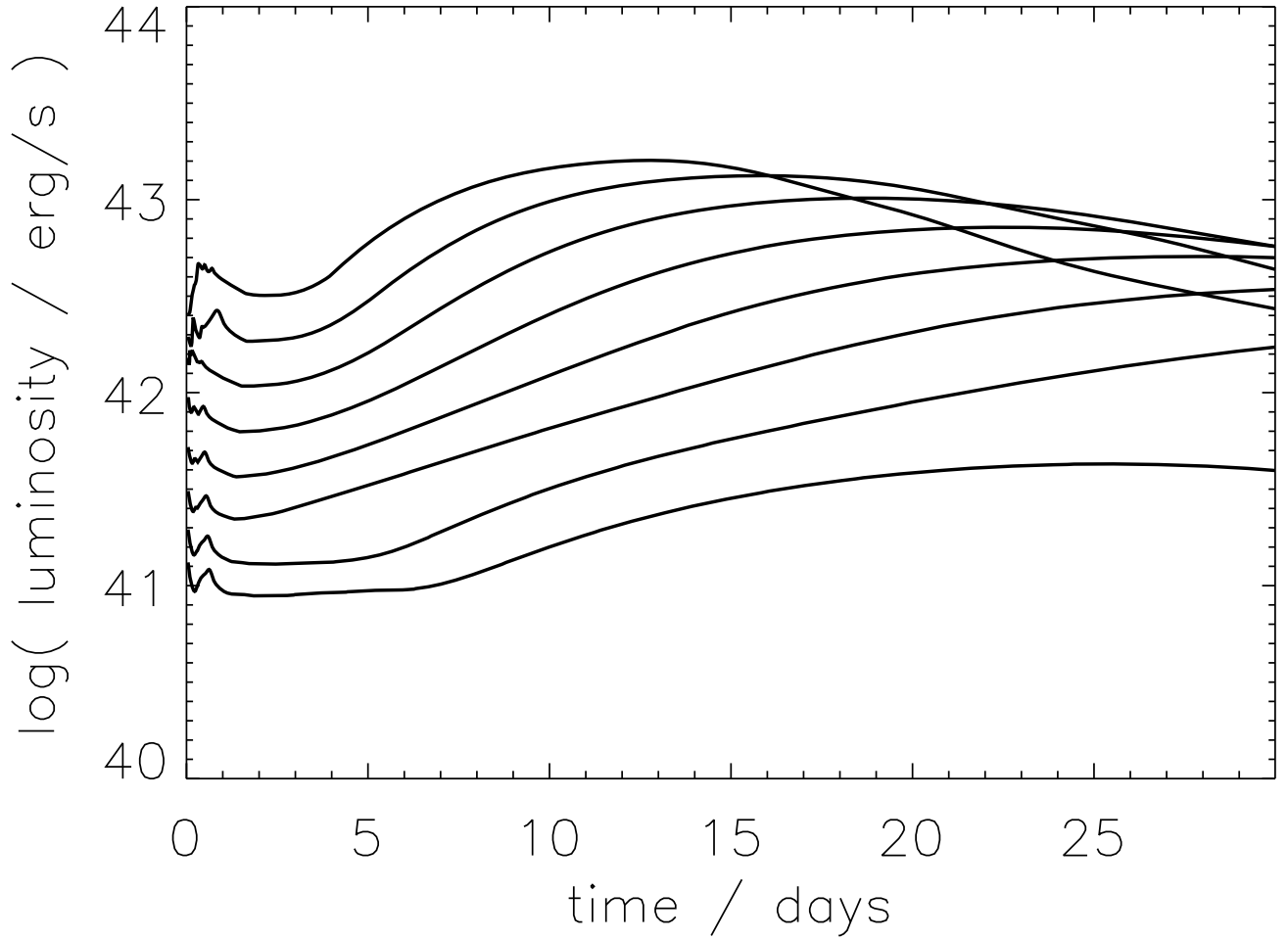}
\caption{Light curves for ``moderately'' mixed models derived from the
high mass progenitor ($15\,\Msun$) and the large piston mass
($2.75\,\Msun$).  The explosions were calculated in spherical symmetry
with kinetic energies at infinity of 160, 80, 40, 20, 10, 5, 2.5, and
1.25$\,\Foe$ and the light curves include just the mass of \I{56}{Ni}
produced explosively in the model (\Tab{mods}).  Note the pre-maximum
non-monotonic evolution of the most energetic models resulting from
helium and heavy element recombination.  For the most energetic
explosions the initial rise for the first few hours is a numerical
artifact of removing zones moving faster than
$1/3\,\cL$. \lFig{lite15}}
\end{figure}

\clearpage
\begin{figure} \centering
\includegraphics[bb=100 20 573 769,clip,angle=90,width=0.525\textwidth]{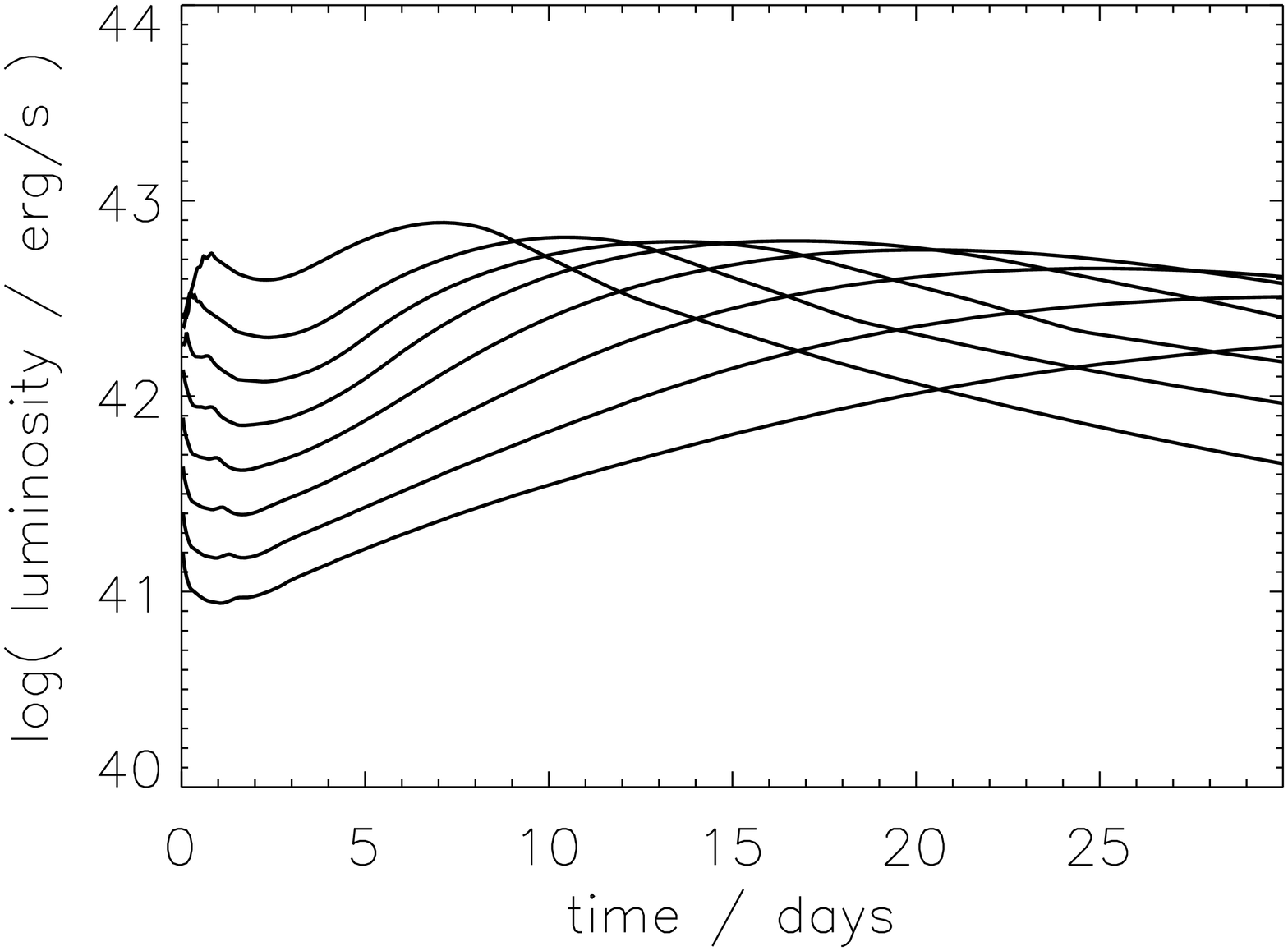}
\includegraphics[bb=100 20 573 670,clip,angle=90,width=0.455\textwidth]{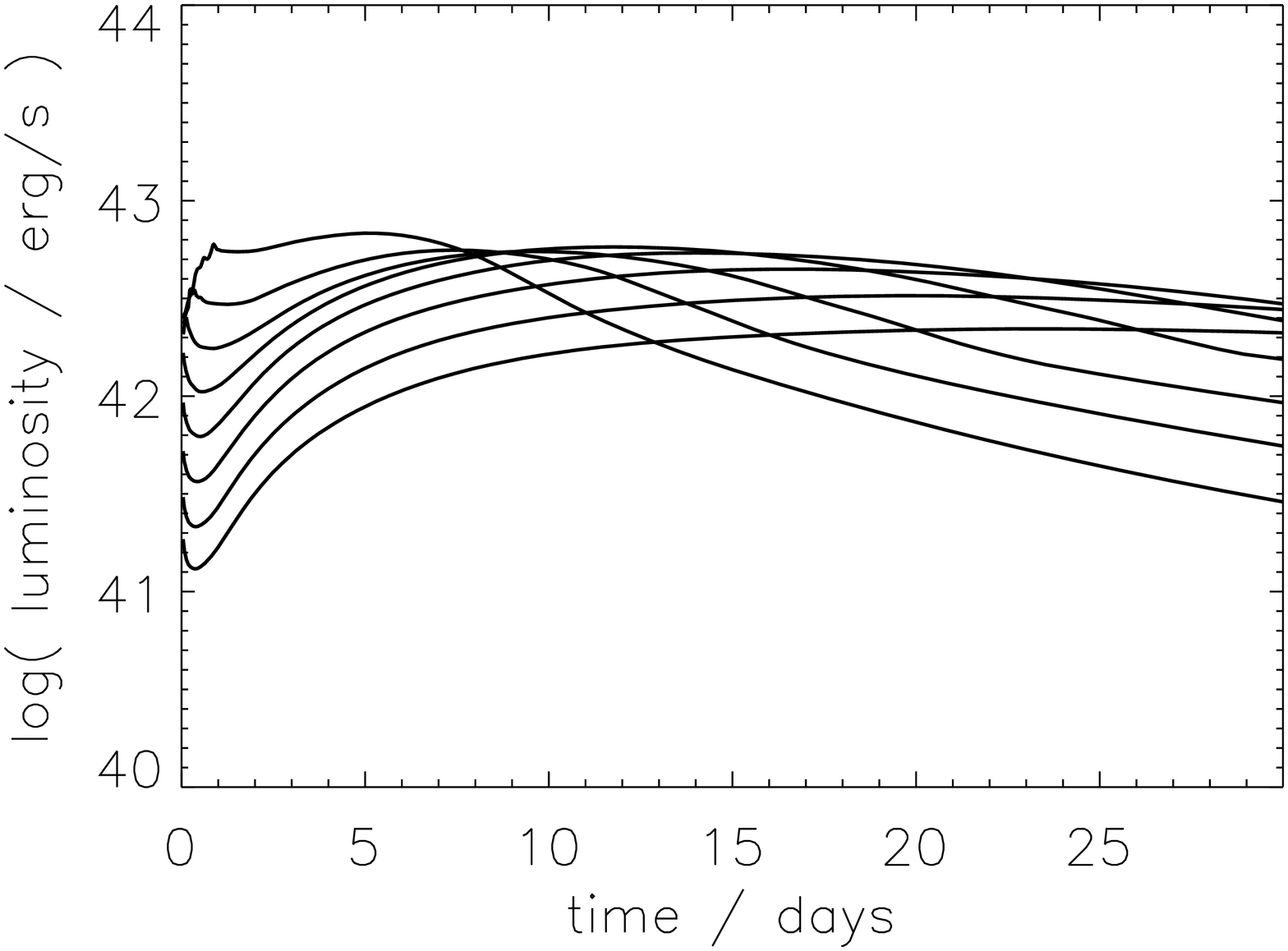}
\includegraphics[bb=021 20 573 769,clip,angle=90,width=0.525\textwidth]{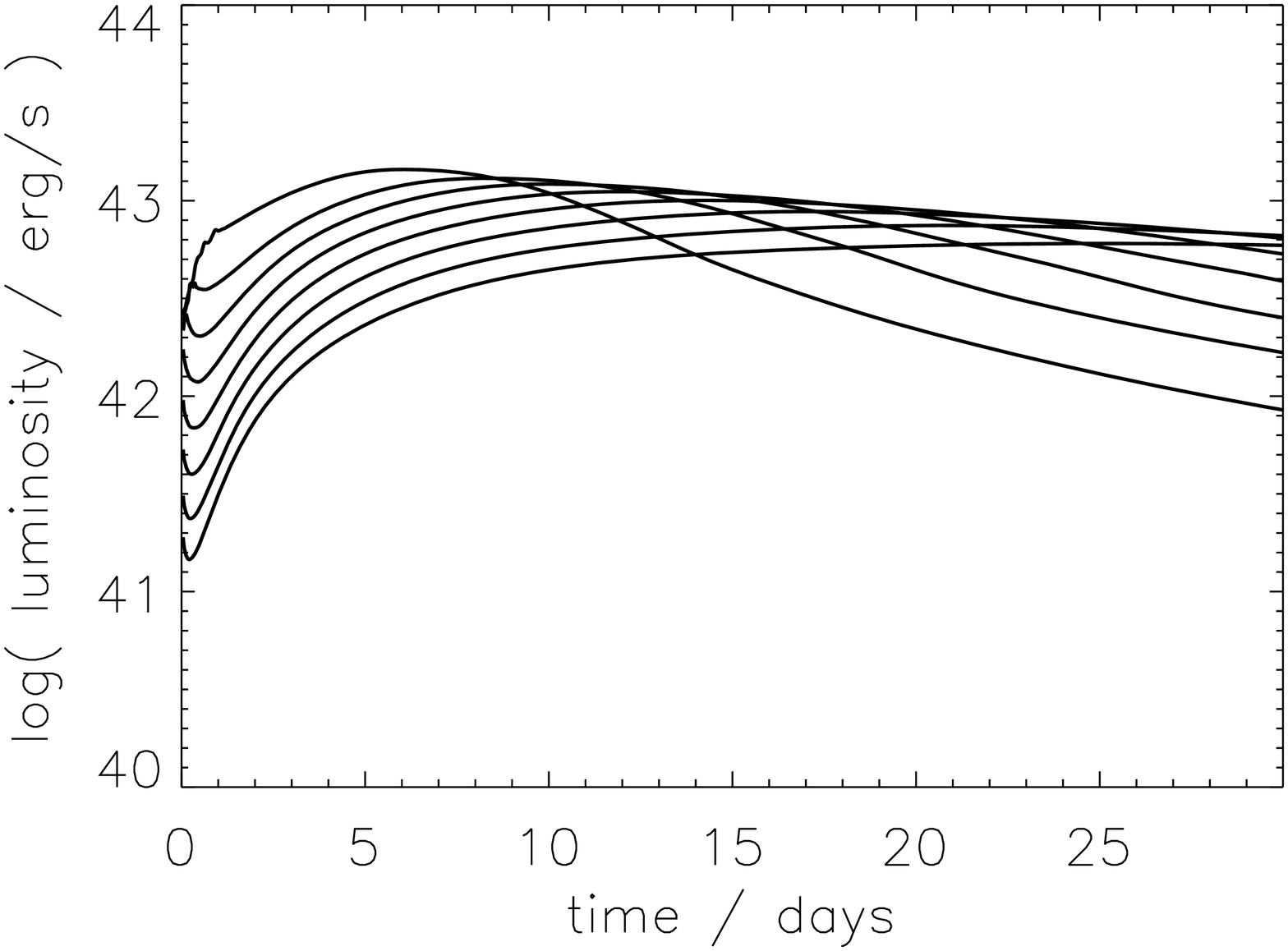}
\includegraphics[bb=021 20 573 670,clip,angle=90,width=0.455\textwidth]{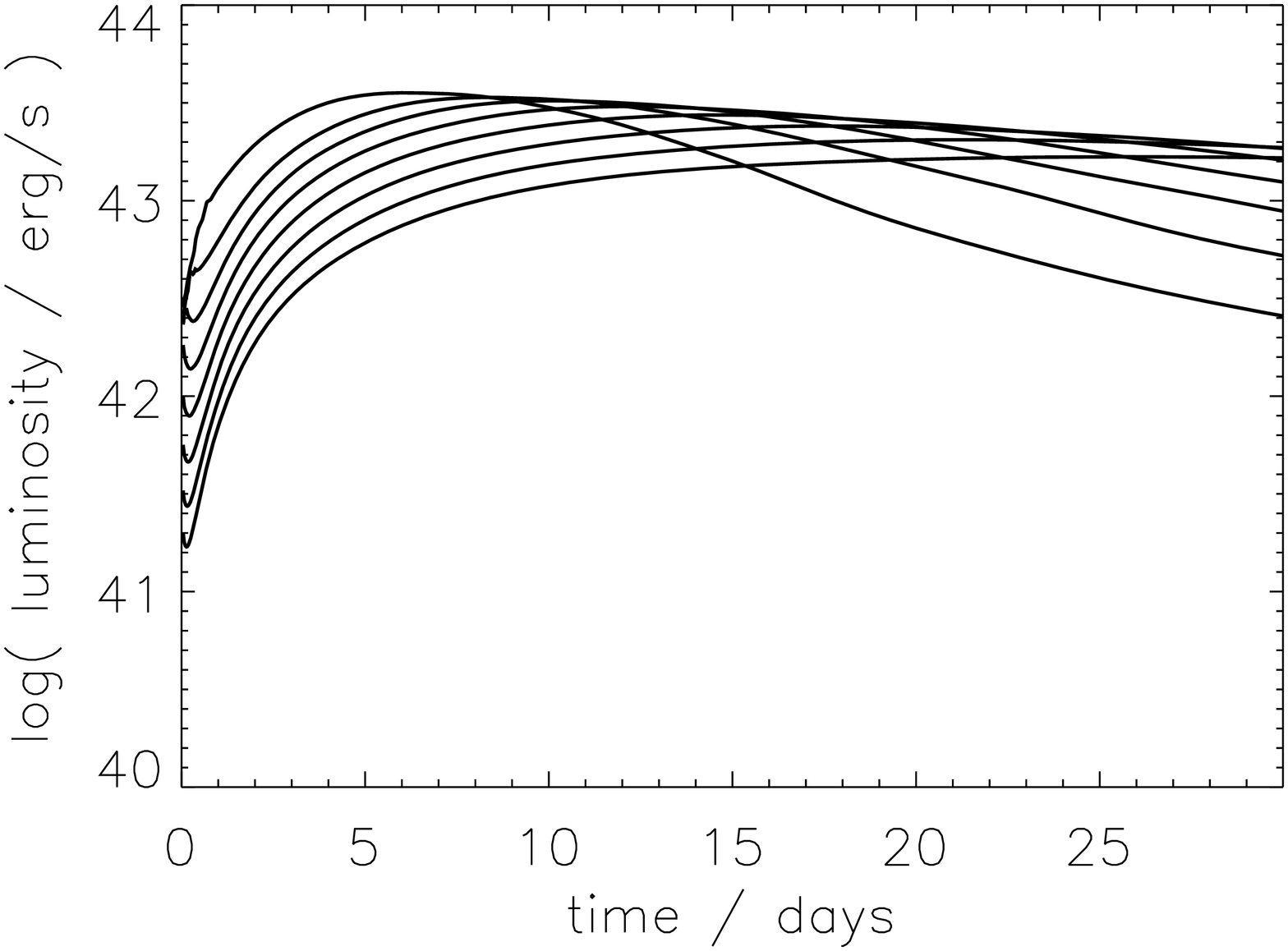}
\caption{Light curves from the models for the low mass progenitor
($8.4\,\Msun$). Calculated in spherical symmetry with kinetic energies
at infinity of 160, 80, 40, 20, 10, 5, 2.5, and 1.25$\,\Foe$. See also
\Fig{lite15}.  \emph{Top Left:} Moderately mixed with the \I{56}{Ni}
yields in \Tab{mods}.  \emph{Top Right:} Same models as top left, but
completely mixed.  \emph{Bottom Left:} Completely mixed models with a
normalized abundance of \I{56}{Ni} of $0.5\,\Msun$ in all cases. The
peaks of the light curves are brighter owing to the increased mass of
\I{56}{Ni} and the pre-maximum variability is less apparent.
\emph{Bottom Right:} Same mixed calculations normalized to a
\I{56}{Ni} mass of $1.5\,\Msun$, as is appropriate for the
high-velocity ejecta.  \lFig{lite8}}
\end{figure}

\clearpage
\begin{figure} \centering
\includegraphics[angle=90,height=0.305\textheight]{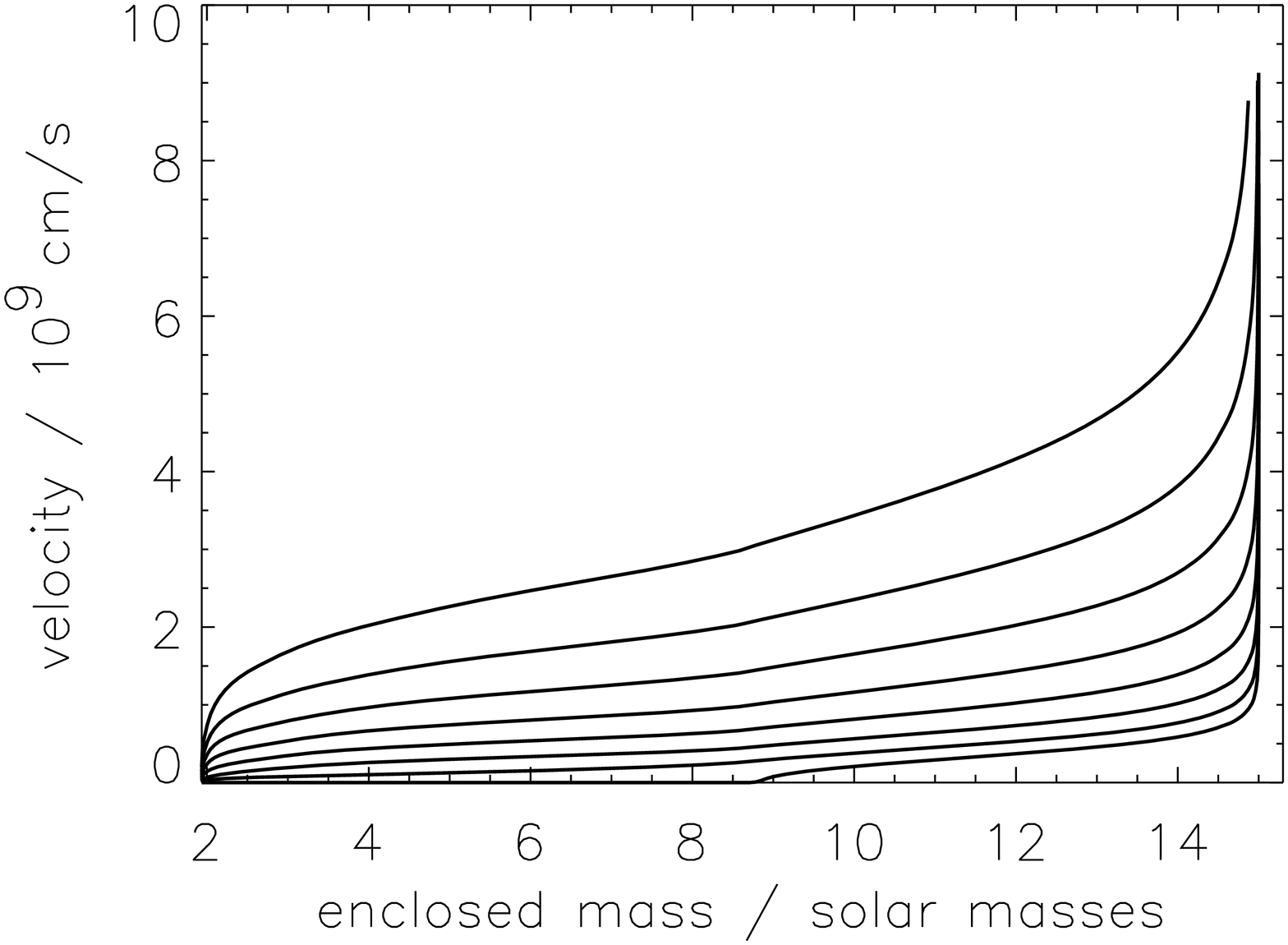}
\includegraphics[bb=21 20 573 660,clip,angle=90,height=0.305\textheight]{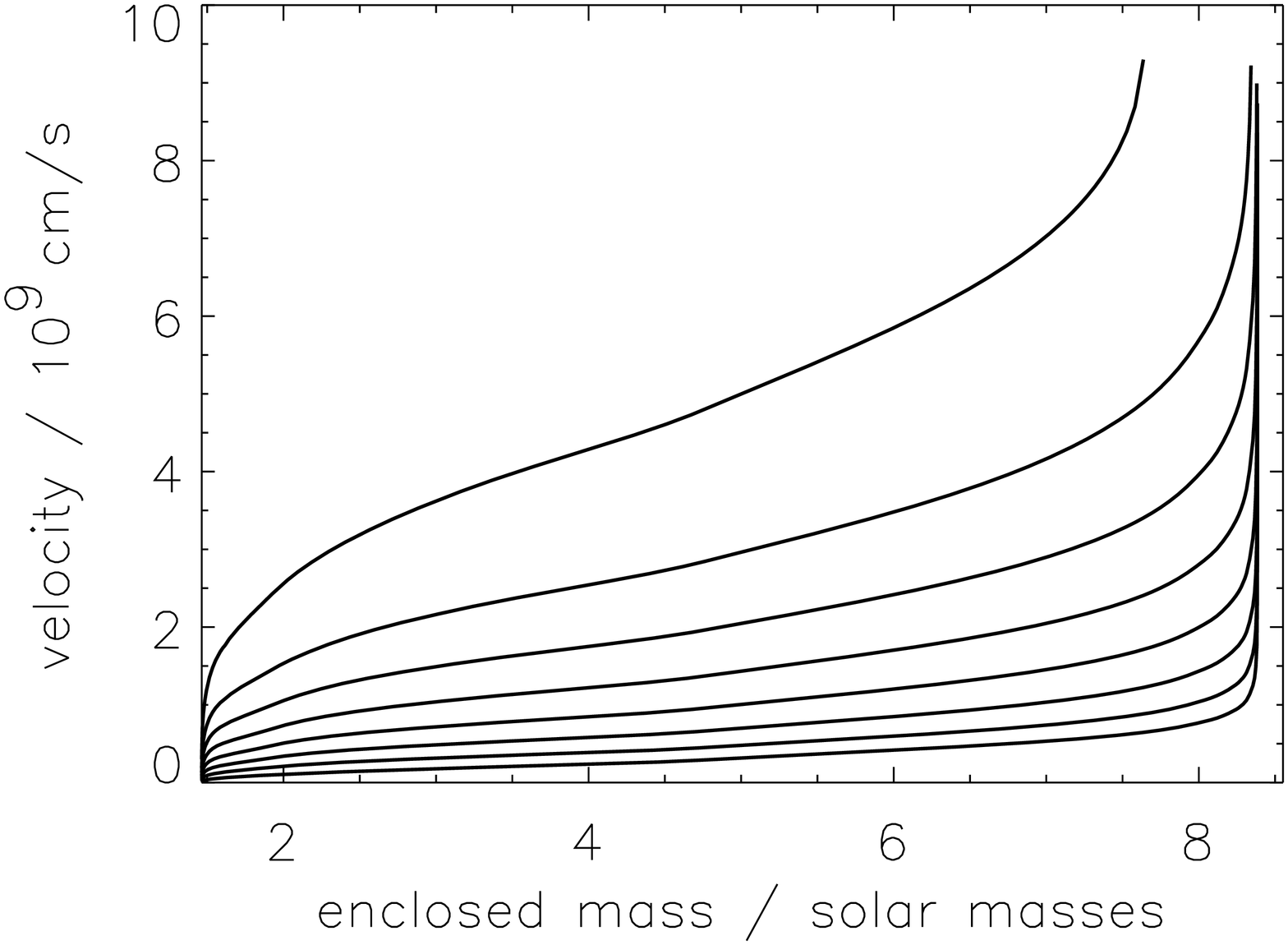}
\caption{\emph{Left:} Final velocity as a function of mass for a
piston location of $1.93\,\Msun$ in the $15\,\Msun$ model.  Moderate
mixing and only the \I{56}{Ni} produced intrinsically by the explosion
were used.  Note fallback of $\sim9\,\Msun$ for the lowest explosion
energy. \emph{Right:} Same plot, but for the $8.38\,\Msun$
model\lFig{vfinal}}
\end{figure}

\clearpage
\begin{figure} \centering
\includegraphics[angle=90,height=0.307\textheight]{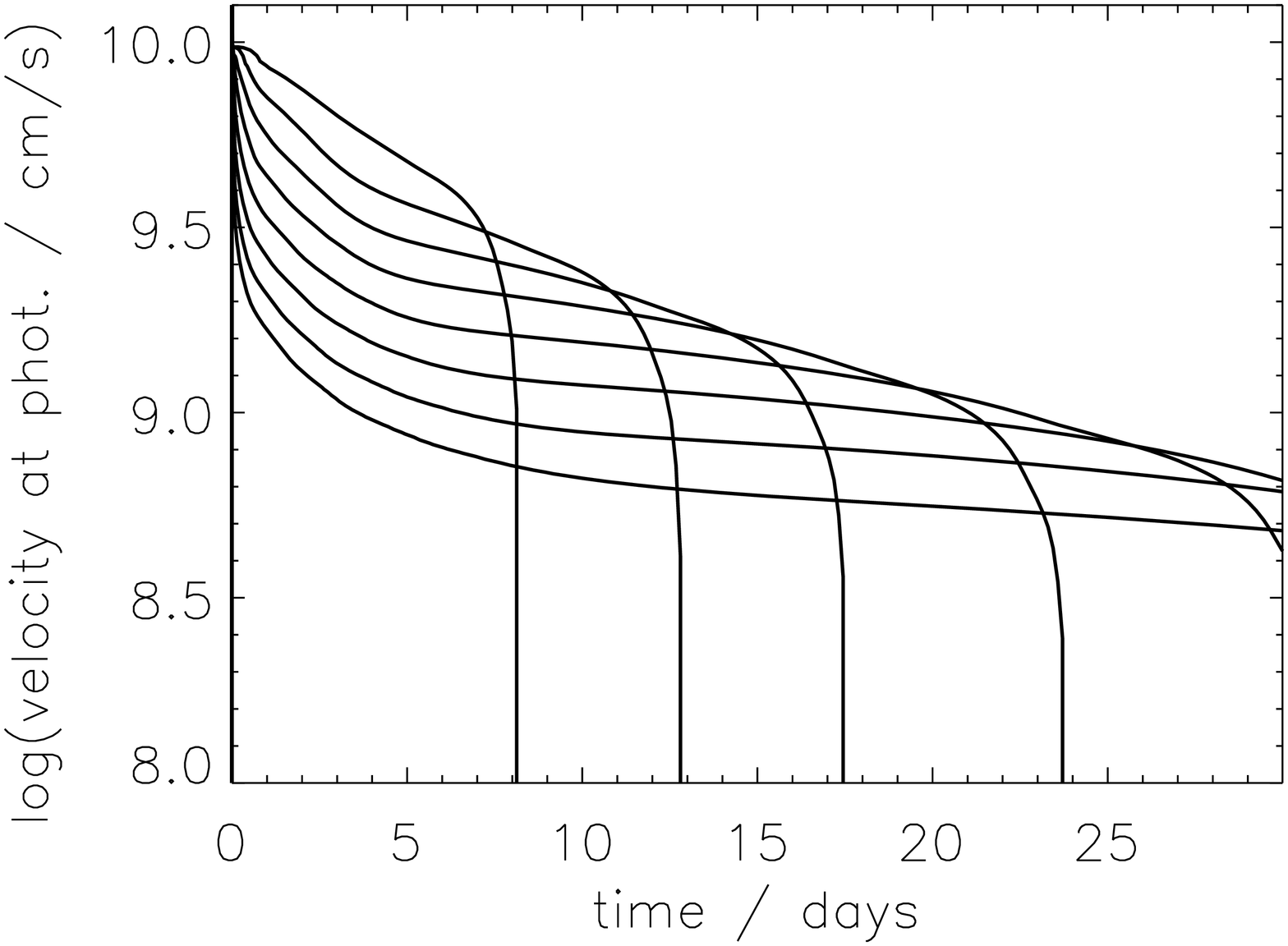}
\includegraphics[bb=21 20 573 645,clip,angle=90,height=0.307\textheight]{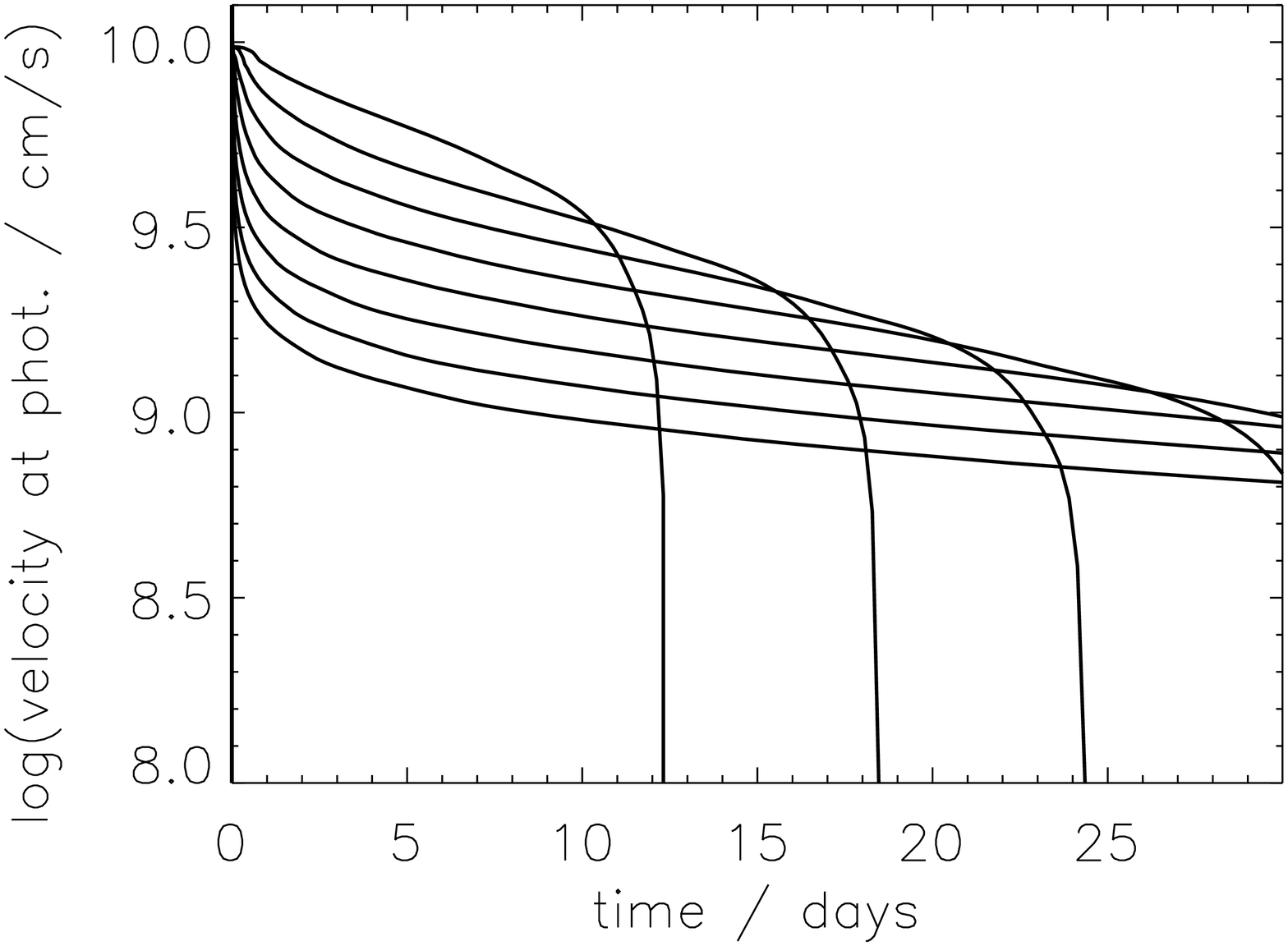}
\caption{\emph{Left:} Velocity at the photosphere for the 160, 80, 40,
20, 10, 5, 2.5, and 1.25$\,\Foe$ explosions of the $8.38\,\Msun$ star
with moderate mixing and only its intrinsic \I{56}{Ni} production.
Note that the velocity drops to zero and becomes undefined thereafter
when the photosphere reaches the center of the ejecta. (A line of
sight through the center would have an optical thickness of $4/3$ at
this point.) Also note that zones moving faster than $\Ep{10}\,\cms$
have been cut off from the calculations; this affects the velocity at
the photosphere for about half a day for the most energetic explosion,
and increasingly earlier times for lower energies. \emph{Right:} Same
plot, but for ejecta whose \I{56}{Ni} mass has been increased to
$1.5\,\Msun$ and the ejecta were ``thoroughly mixed''.  Extra heating
by the increased \I{56}{Ni} abundance at large radii maintains an
extended photosphere longer.  \lFig{vphot}}
\end{figure}

\clearpage
\begin{figure} \centering
\includegraphics[angle=90,height=0.6\textheight]{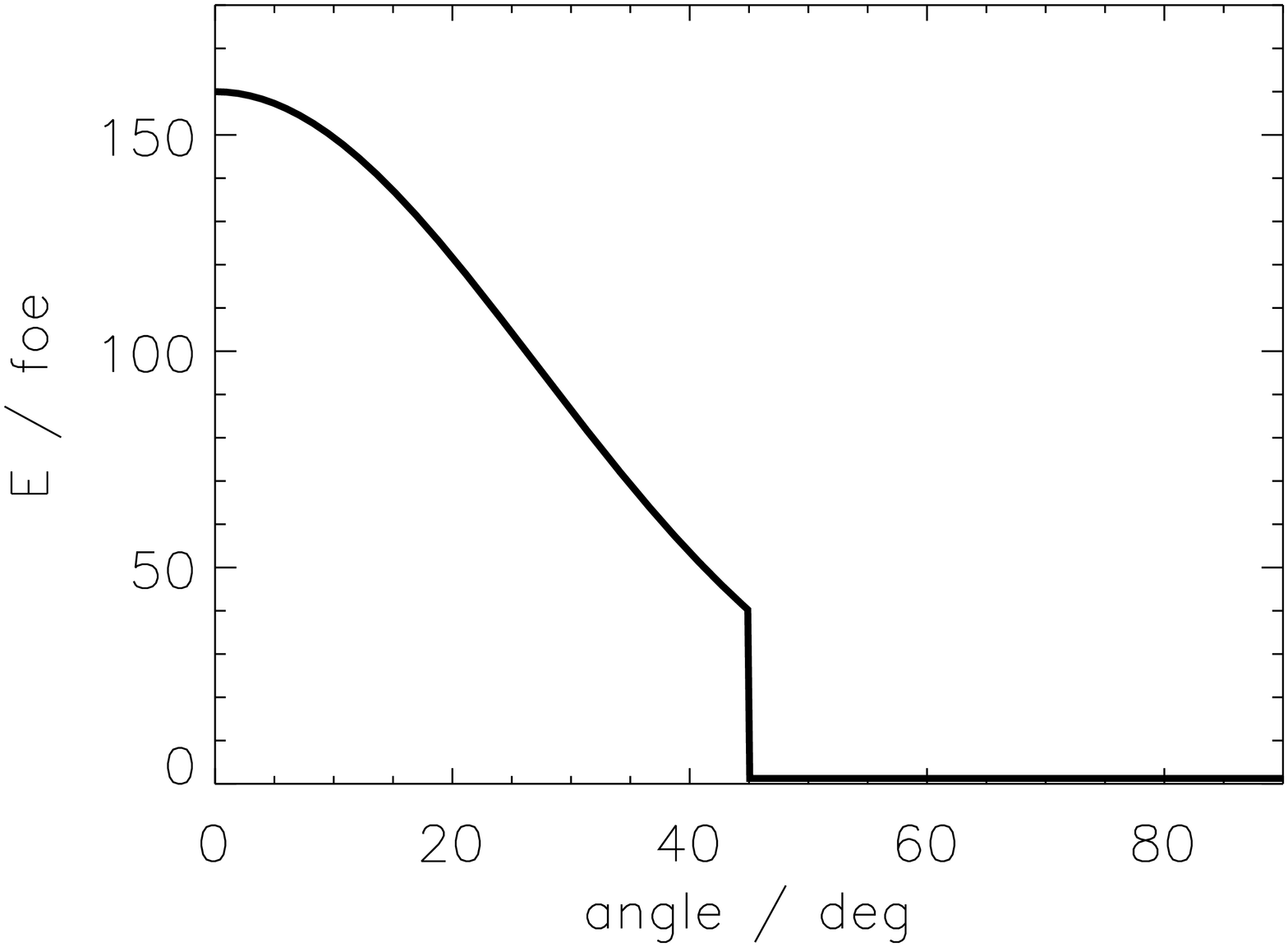}
\caption{Contributions to the composite model. The figure gives the
equivalent isotropic energy along a given angle as a function of
angle.  For angles less than $45\Deg$ the energy varies from
$160\,\Foe$ to $40\,\Foe$ and models normalized to $1.5\,\Msun$ of
\I{56}{Ni} were used, being ``thorough'' mixed.  For angles greater
than $45\Deg$ the explosion energy is $1.25\,\Foe$ weak; only the
intrinsic \I{56}{Ni} production (\Tab{proj2}) and mild mixing was
employed.\lFig{eiso}}
\end{figure}

\clearpage
\begin{figure} \centering
\includegraphics[angle=0,bb=23 295 568 548,width=\columnwidth]{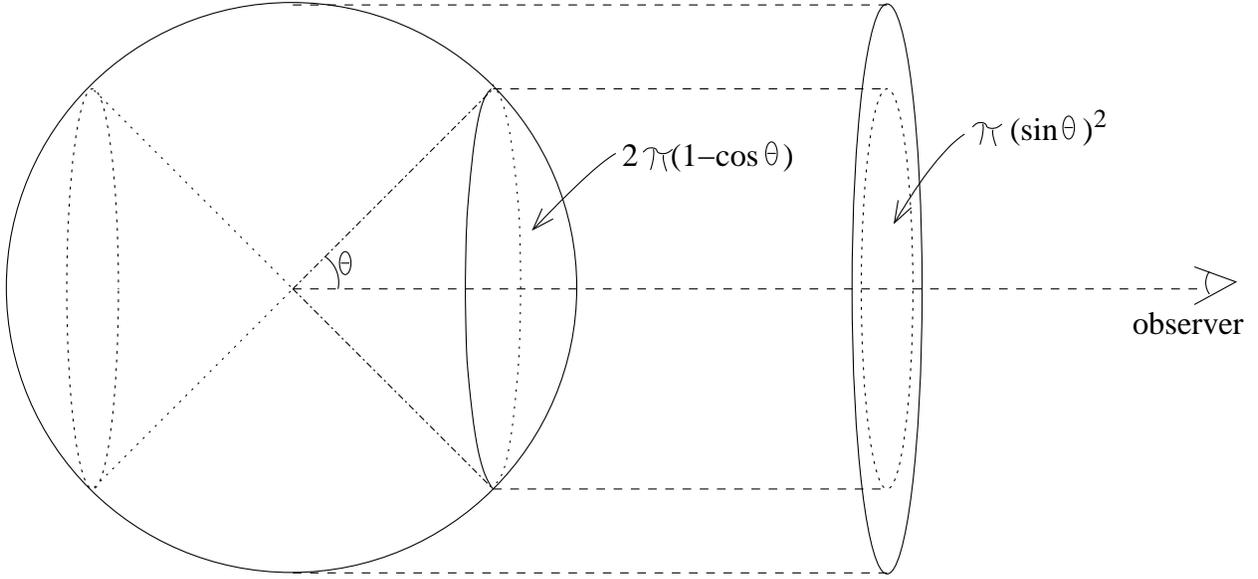}
\caption{Decomposition of the light curve.  Assuming, a (bipolar) jet of
opening angle $2\theta$ (e.g., $\theta=45\Deg$ as depicted above),
each polar cap comprises a solid angle of
$2\pi\left(1-\cos\theta\right)$, i.e., the bipolar jet flows out at a
fraction $1-\cos\theta$ of the total solid angle of $4\pi$.  This
means $1-\cos\theta$ of the isotropic explosion contributes to such an
explosion and its energy.  However, an observer looking along the axis
of the jet sees the projected surface area
$\pi\left(\sin\theta\right)^2$ out of $2\pi$ for the full circle, the
full projected sphere, i.e., a fraction $\sin^2\theta$ of the
isotropic light curves contributes to apparent light curve for a such
positioned observer.  Note that for $0\Deg<\theta<90\Deg$:
$\sin^2\theta > 1-\cos\theta$, i.e., the observer along the axis is
favored to see an apparently brighter light curve than would
correspond to the total energy of a composed light curve (assuming the
explosion is stronger in polar direction than in equatorial
direction).  This is because material at high colatitude, $\theta$,
appears under a high inclination angle.  \lFig{slice}}
\end{figure}

\clearpage
\begin{figure} \centering
\includegraphics[angle=90,height=0.6\textheight]{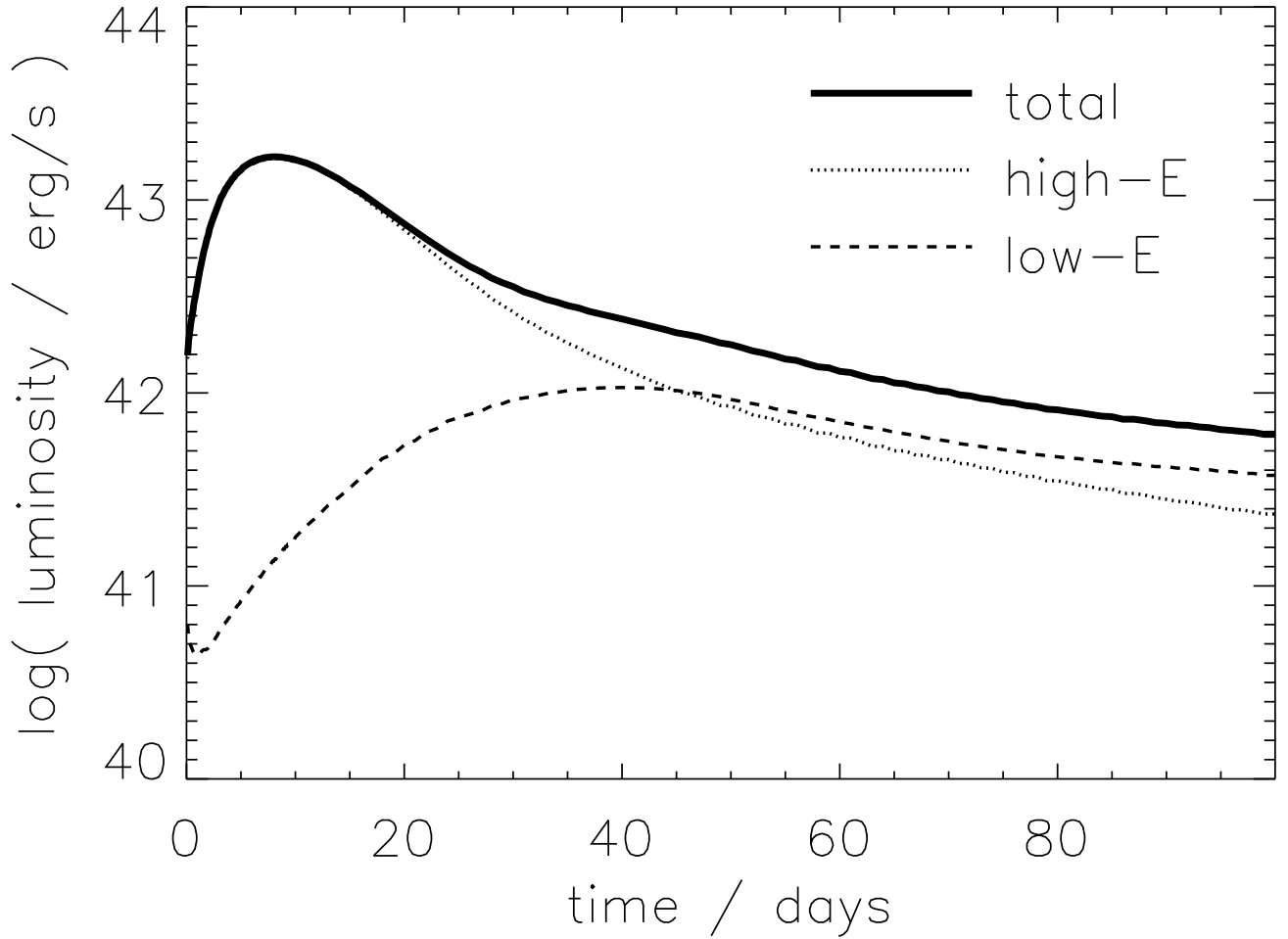}
\caption{Composite light curve resulting from a combination of
well-mixed models (\Fig{lite8}) having a distribution of kinetic
energies as given in \Fig{eiso}.  For polar angles less than $45\Deg$
the three highest energy models in \Fig{lite8} were employed which
each have an effective \I{56}{Ni} mass of $1.5\,\Msun$.  Together,
with appropriate angular weighting (see text), these give the curve
labeled ``high-E''. For angles greater than $45\Deg$ the model with
energy $1.25\,\Foe$ in \Fig{lite8} was used. The \I{56}{Ni} mass in
this component was left at $0.15\,\Msun$. This curve, with appropriate
weighting, is labeled ``low-E''. The dark solid line is the total. The
total energy in the composite is $26\,\Foe$ and the total mass of
\I{56}{Ni} is $0.55\,\Msun$. \lFig{composite}}
\end{figure}


\begin{thebibliography}

\bibitem[Braun(1997)]{Bra97}
Braun, H. 1997, PhD thesis, Ludwig-Maximilians-Universit\"at M\"unchen

\bibitem[Buchmann(1996)]{Buc96} 
Buchmann, L.\ 1996, \apjl, 468, L127

\bibitem[Chugai(2000)]{Chu00} 
Chugai, N.~N.\ 2000, Astronomy Letters, 26, 797

\bibitem[Frail et al.(2001)]{Fra01}
Frail, D., Kulkarni, S. R., Sari, R., Djorgovski, S. G., Bloom, J. S.,
Galama, T. J. et al. 2001, \apjl, 562, L55

\bibitem[Galama et al.(1998)]{Gal98}
Galama, T. J., et al. 1998, Nature, 395, 670 

\bibitem[Hamann \& Koesterke(1998)]{HK98}
Hamann, W.-R., Koesterke, L. 1998, \aap, 335, 1003

\bibitem[Heger, Langer, \& Woosley(2000)]{Heg00} 
Heger, A., Langer, N., \& Woosley, S.~E.\ 2000, \apj, 528, 368

\bibitem[Hjorth et al.(2003)]{Hjo03}
Hjorth, J., Sollerman, J., M\"oller, P., Fynbo, J. P. U., Woosley,
S. E., Kouveliotou, C., et al. 2003, Nature, 423, 847

\bibitem[H\"oflich, Wheeler, \& Wang(1999)]{Hof99}
H\" oflich, P., Wheeler, J.~C., \& Wang, L.\ 1999, \apj, 521, 179

\bibitem[Iwamoto et al.(1998)]{Iwa98} 
Iwamoto, K.Mazzali, P. A., Nomoto, K., Umeda, H., Nakamura, T., Patat,
F., ~et al.\ 1998, \nat, 395, 672

\bibitem[Leibundgut(1994)]{Lei94} 
Leibundgut, B.\ 1994, in {\sl Circumstellar Media in Late Stages of
Stellar Evolution}, eds. R.E.S. Clegg, I.R. Stevens \& W.P.S. Meikle,
(Cambridge University Press),100

\bibitem[Li \& Chevalier(1999)]{Li99} 
Li, Z.~\& Chevalier, R.~A.\ 1999, \apj, 526, 716

\bibitem[MacFadyen \& Woosley(1999)]{Mac99}
MacFadyen, A. I., \& Woosley, S. E. 1999, \apj, 524, 262

\bibitem[MacFadyen(2002)]{Mac02}
MacFadyen, A. I. 2002, in {\sl From Twilight to Highlight: The Physics
of Supernovae}, eds. W. Hillebrandt \& B. Leibundgut, ESO Astrophysics
Symposia, p. 97.

\bibitem[Maeda et al.(2002)]{Mae02} 
Maeda, K., Nakamura, T., Nomoto, K., Mazzali, P.~A., Patat, F., \&
Hachisu, I.\ 2002, \apj, 565, 405

\bibitem[Maeda et al.(2003)]{Mae03} 
Maeda, K., Mazzali, P.~A., Deng, J., Nomoto, K., Yoshii, Y., Tomita,
H., \& Kobayashi, Y.\ 2003, \apj, 593, 931

\bibitem[Mazzali et al.(2001)]{Maz01} 
Mazzali, P.~A., Nomoto, K., Patat, F., \& Maeda, K.\ 2001, \apj, 559,
1047

\bibitem[Nakamura et al.(2001)]{Nak01}
Nakamura, T., Mazzali, P.~A., Nomoto, K., \& Iwamoto, K.\ 2001, \apj,
550, 991

\bibitem[Pruet, Woosley, \& Hoffman(2003)]{Pru03} 
Pruet, J., Woosley, S.~E., \& Hoffman, R.~D.\ 2003, \apj, 586, 1254

\bibitem[Sollerman et al.(2002)]{Sol02} 
Sollerman, J.~et al.\ 2002, \aap, 386, 944

\bibitem[Stanek et al.(2003)]{Sta03} 
Stanek, K.~Z., Matheson, T., Garnavich, P. M., Martini, P., Berlind, P.,
Caldwell, N., ~et al.\ 2003, \apjl, 591, L17

\bibitem[Weaver, Zimmerman, \& Woosley(1978)]{Wea78} 
Weaver, T.~A., Zimmerman, G.~B., \& Woosley, S.~E.\ 1978, \apj, 225,
1021

\bibitem[Wheeler, Yi, H{\" o}flich, \& Wang(2000)]{Whe00} 
Wheeler, J.~C., Yi, I., H{\" o}flich, P., \& Wang, L.\ 2000, \apj,
537, 810

\bibitem[Willingale et al.(2003)]{Wil03}
Willingale, R., Osborne, J. P., O'Brian, P. T., Ward, M. J., Levan,
A., \& Page, R. L. 2003, submitted to MNRAS, astroph-0307561.

\bibitem[Woosley, Eastman, \& Schmidt(1999)]{Woo99} 
Woosley, S.~E., Eastman, R.~G., \& Schmidt, B.~P.\ 1999, \apj, 516,
788

\bibitem[Zhang, Woosley, \& MacFadyen(2002)]{Zha02}
Zhang, W., Woosley, S. E., \& MacFadyen, A. I. 2002, \apj, in press,
astro-ph/0207436

\end{thebibliography}
\end{document}